\documentclass[11pt]{article}

\usepackage{ifthen}
\usepackage{latexsym}
\usepackage{amsfonts}
\usepackage{amsmath}
\usepackage{amssymb}
\usepackage{pifont}

\newcommand{\hugeDebug}{false}
\ifthenelse{\equal{\hugeDebug}{false}}{
\newcommand{\normalspacing}{\singlespacing}
\pagestyle{plain}
\setlength{\oddsidemargin}{0.25in}
\setlength{\evensidemargin}{\oddsidemargin}
\setlength{\textwidth}{6in}
\setlength{\textheight}{8in}
\setlength{\topmargin}{-0.0in}
}
{
\newcommand{\normalspacing}{\niceninespacing}
\pagestyle{headings}
 \setlength{\oddsidemargin}{-0.4in}
 \setlength{\evensidemargin}{\oddsidemargin}
 \setlength{\textwidth}{5.3in}
 \setlength{\textheight}{6in}
 \setlength{\textheight}{6.25in}
 \setlength{\topmargin}{-0.0in}
}

\newlength{\filength}
\newsavebox{\gcbox}
\sbox{\gcbox}{\framebox[\filength]{\rule{0ex}{2ex}}}

\newcommand{\sparse}{{{\rm SPARSE}}}

  \newtheorem{theorem}{Theorem}[section]

\newcommand\qedblob{\ding{113}}
\def\literalqed{{\ \nolinebreak\hfill\mbox{\qedblob\quad}}}

\showhyphens{}

\newcount\hour  \newcount\minutes  \hour=\time  \divide\hour by 60
\minutes=\hour  \multiply\minutes by -60  \advance\minutes by \time
\def\mmmddyyyy{\ifcase\month\or Jan\or Feb\or Mar\or Apr\or May\or Jun\or Jul\or
  Aug\or Sep\or Oct\or Nov\or Dec\fi \space\number\day, \number\year}
\def\hhmm{\ifnum\hour<10 0\fi\number\hour :%
  \ifnum\minutes<10 0\fi\number\minutes}

\makeatletter
\def\@citex[#1]#2{\if@filesw\immediate\write\@auxout{\string\citation{#2}}\fi
  \def\@citea{}\@cite{\@for\@citeb:=#2\do
    {\@citea\def\@citea{,\linebreak[0]}\@ifundefined
       {b@\@citeb}{{\bf ?}\@warning
       {Citation `\@citeb' on page \thepage \space undefined}}%
\hbox{\csname b@\@citeb\endcsname}}}{#1}}
\newcommand{\singlespacing}{\let\CS=
\@currsize\renewcommand{\baselinestretch}{1}\tiny\CS}
\newcommand{\singlespacingplus}{\let\CS=
\@currsize\renewcommand{\baselinestretch}{1.25}\tiny\CS}
\newcommand{\doublespacing}{\let\CS=
\@currsize\renewcommand{\baselinestretch}{1.75}\tiny\CS}
\newcommand{\extradoublespacing}{\let\CS=
\@currsize\renewcommand{\baselinestretch}{1.9}\tiny\CS}
\newcommand{\nicenicespacing}{\let\CS=
\@currsize\renewcommand{\baselinestretch}{1.9}\tiny\CS}
\newcommand{\draftspacing}{\let\CS=
\@currsize\renewcommand{\baselinestretch}{2.0}\tiny\CS}
\newcommand{\hugedraftspacing}{\let\CS=
\@currsize\renewcommand{\baselinestretch}{2.4}\tiny\CS}
\newcommand{\niceonespacing}{\let\CS=\@currsize\renewcommand{\baselinestretch}{1.1}\tiny\CS}
\newcommand{\nicetwospacing}{\let\CS=\@currsize\renewcommand{\baselinestretch}{1.2}\tiny\CS}
\newcommand{\nicethreespacing}{\let\CS=\@currsize\renewcommand{\baselinestretch}{1.3}\tiny\CS}
\newcommand{\singlespacingplusplus}{\let\CS=\@currsize\renewcommand{\baselinestretch}{1.35}\tiny\CS}
\newcommand{\nicefourspacing}{\let\CS=\@currsize\renewcommand{\baselinestretch}{1.4}\tiny\CS}
\newcommand{\nicefivespacing}{\let\CS=\@currsize\renewcommand{\baselinestretch}{1.5}\tiny\CS}
\newcommand{\nicesixspacing}{\let\CS=\@currsize\renewcommand{\baselinestretch}{1.6}\tiny\CS}
\newcommand{\nicesevenspacing}{\let\CS=\@currsize\renewcommand{\baselinestretch}{1.7}\tiny\CS}
\newcommand{\niceeightspacing}{\let\CS=\@currsize\renewcommand{\baselinestretch}{1.8}\tiny\CS}
\newcommand{\niceninespacing}{\let\CS=\@currsize\renewcommand{\baselinestretch}{1.9}\tiny\CS}
\makeatother

\normalspacing



\renewcommand{\exp}{{\rm EXP}}
\newcommand{\NE}{{\rm NE}}
\renewcommand{\ne}{{\rm NE}}
\newcommand{\nexp}{{\rm NEXP}}
\newcommand{\p}{{\rm P}}

\newcommand{\ntime}{{\rm NTIME}}

\newcommand{\np}{{\rm NP}}

\newcommand{\optp}{{\rm OptP}}

\newcommand{\conp}{{\rm coNP}}

\newcommand{\pnexp}{\ensuremath{\p^\nexp}}

\newcommand{\acc}{{\rm ACC}}
\newcommand{\acczero}{\ensuremath{\acc^0}}
\newcommand{\accone}{\ensuremath{\acc^1}}
\newcommand{\acck}{\ensuremath{\acc^k}}

\newcommand{\poly}{\ensuremath{{\rm poly}}}

\newcommand{\pne}{\ensuremath{\p^\ne}}
\newcommand{\npne}{\ensuremath{\np^\ne}}
\newcommand{\pnpne}{\ensuremath{\p^{\np^\ne}}}
\newcommand{\npnpne}{\ensuremath{\np^{\np^\ne}}}

\newcommand{\expnp}{\ensuremath{\exp^\np }}

\newcommand{\calc}{\ensuremath{{\cal C}}}
\newcommand{\cald}{\ensuremath{{\cal D}}}

\newcommand{\bigo}{{\protect\cal O}}

\newcommand{\fp}{{\rm FP}}

\title{A Note on Nonuniform versus Uniform\\ACC$^k$ Circuits for NE\thanks{Also appears as URCS-TR-2010-964}}
\author{
Lane A. Hemaspaandra\thanks{URL: \mbox{\tt{}www.cs.rochester.edu/u/lane}.
Supported 
in part by
grant
NSF-CCF-0915792 and a 
Friedrich Wilhelm Bessel Research Award.}
\\Department of Computer Science \\
University of Rochester \\
Rochester, NY 14627, USA
}

\date{%
December 2, 2010; revised December 3, 2010}

\begin{document}

\sloppy

\maketitle

\begin{abstract}
We note that for each $k \in \{0,1,2,\ldots\}$ the following 
holds:
NE has (nonuniform) $\acck$ circuits if and only if 
NE has $\pne$-uniform $\acck$ circuits.
And we mention how to get analogous results for 
other circuit and complexity classes.
\end{abstract}

\section{Introduction and Result}
Ryan Williams recently announced the breakthrough advance that some $\NE$
sets lack $\acczero$ circuits~\cite{wil:unpub:nonuniform-acc}.  His result is for the
extremely strong case of defeating even nonuniform $\acczero$ circuits.

Is there some on-the-surface-weaker claim---about 
defeating \emph{uniform} $\acczero$ circuits---that 
is equivalent to this?  This brief note looks at that question, for the
case of each $\acck$.  What we observe is the following.

\begin{theorem}\label{t:main}
For each $k \in \{0,1,2,\ldots\}$ the following 
holds:
$\NE$ has (nonuniform) $\acck$ circuits if and only if 
$\NE$ has $\pne$-uniform $\acck$ circuits.
\end{theorem}

The immediate natural question to ask is: Why should one care about
this?  After all, for $k=0$ Williams's result already handles the most
challenging case, nonuniform $\acczero$, and the above result simply let
one conclude, from his result, a far weaker result.  However, there
are two related reasons why one should care about the above theorem.
First, regarding $k=0$, the goal of the above result isn't to extend
Williams's result, but rather is to understand it better, and in
particular, to understand what seemingly weaker uniform
result---which, we should stress, no one ever obtained---would have
implied Williams's nonuniform result.  
That is, instead of pole
vaulting over a 10-meter-high bar and shattering the world record,
Williams could have indirectly achieved the same strength-of-result
by pole vaulting over a
bar that was merely 9.99 meters high.  Second and more important, 
for $k>0$, the above
result potentially puts in place a very slightly lower bar for whoever
tries to show that, for example, nonuniform $\accone$ circuits cannot
handle all of $\NE$.  Instead of trying to default nonuniform
circuits, that researcher 
need only (although that is a huge ``only'') defeat
$\pne$-uniform circuits.  And although the above theorem says that
that is logically the same as defeating nonuniform $\accone$ circuits,
it is quite possible that one can more easily (although that is
unlikely to be an easy ``easily'') argue regarding the limitations of
circuits of limited-complexity uniformity than one can about
nonuniform circuits.  Although we won't repeat here the history and
background that are well-covered in the paper of Williams, it is worth
mentioning that some of the results that proceeded Williams's work
centrally used the uniformity of the classes being defeated (see the
work of Allender and
Gore~\cite{all-gor:j:permanent,all:j:permanent-threshold}).

The proof of the theorem is quite brief.  One just takes the brute-force 
bound (please see 
footnote~\ref{foot:Hopcroft}
for history and for credit/relation to Hopcroft)
one gets from guessing and checking 
circuits,\footnote{\label{foot:Hopcroft}Guessing and checking is an often helpful approach 
in complexity theory.  It is employed in Williams's 
proof~\cite{wil:unpub:nonuniform-acc},
and it has a long history, e.g., one can find it in 
Hopcroft's alternate proof of the 
Karp--Lipton Theorem~(\cite{hop:c:recent}, and as that 
is a conference-length-only paper and may be a bit hard 
to find, please note that a more recent writeup of that 
proof that very explicitly and in detail follows 
the guess-and-check approach of Hopcroft can be 
find as~\cite[Proof of Theorem~1.16]{hem-ogi:b:companion})
and in the work 
of 
Balc{\'{a}}zar, Book, Long, Sch{\"{o}}ning, and	
Selman~\cite{bal-boo-sch:j:sparse,lon-sel:j:sparse}
showing that 
the polynomial hierarchy collapses 
if and only if 
the polynomial hierarchy collapses relative to some sparse set
if and only if 
the polynomial hierarchy collapses relative to every sparse set.

In fact, it is important as credit-where-credit-is-due to mention that
in a very real sense the Hopcroft insight mention above \emph{is} (or
is very close to) the same ``brute-force'' approach we're focusing on
in this note (although the point of this note is mostly to note that
for the case of NE an unexpected simplification occurs, and also to make
explicit how the brute-force approach plays out in a
circuit-uniformity setting; however, in fact, in some ways, the Hopcroft work
is better than brute force---see below---as it uses an 
additional wonderful trick specific to its own setting).

The following rather technical expansion on the comment just made is
addressed only to those who are familiar with the Hopcroft approach.

What we mean when we say that 
the Hopcroft ``guess the sparse set 
(or circuit, as $\p^\sparse = \p/\poly$ and so guessing 
sparse sets and guessing a small general circuit are almost 
the same, in this context)
and check'' proof of the Karp--Lipton
Theorem
really is doing the same thing, or very, very closely to the same thing, 
as 
the brute-force approach mentioned here
is that the only real difference
(when one looks beyond the surface and thanks about the flavor of 
what is going on)
is that in the circuit uniformity case one is \emph{producing} a
circuit, and thus we in our oracle stack of classes have an FP on
bottom, but in contrast Hopcroft's proof can have lots of different
paths that guess good sparse sets (or circuits) and that is no problem
there as they all will do the right thing and there will be at least
one such path; and a second difference is that Hopcroft actually
uses \emph{less} of a stack of quantifier access (even aside from the
extra FP on bottom for the reason just mentioned), because he isn't
merely brute-forcing things, but is using the utterly lovely trick of
the 2-disjunctive-self-reducibility of SAT, which lets him with a
single ``forall''-type oracle call
check the consistency of each internal node of the
self-reducibility tree and also check the leaves and by doing so
already know if the given path has a good sparse set (or circuit), 
and if so then that path then uses that setting's 
own $\np^\np$ to handle the 
first two levels of the target $\np^{\np^\np}$ set while passing
up (along with other things)
the sparse set (or circuit) so that the upper of the two $\np$'s
can use the sparse set (or circuit) 
to handle the top-most $\np$ of the height-3 target set,
and that works perfectly in his setting.   
That latter difference is not a generally available trick, but it works
like magic in his $\calc = \np$ case, 
and actually, in his setting that magic seems needed
to get the collapse from the proof approach
(although historically there already was the quite different
Karp-Lipton proof).}
sees that
one gets a bound at the $\Delta_4$ function-level of the 
so-called strong
exponential hierarchy, and then invokes a result from the 1980s that
shows that the strong exponential hierarchy (whose levels are 
E, NE, $\pne$, $\npne$, $\pnpne$, $\npnpne$,~$\ldots$\footnote{As 
a brief bit of context, taken 
from \cite{hem:j:sky}, we mention that by padding, $\pne = \pnexp$
(recall $\ne$ is $\ntime[2^{\bigo(n)}]$ and 
$\nexp$ is $\ntime[2^{\bigo(n^{\bigo(1)})}]$),
and that $\pne \subseteq \expnp$ but it is not at all clear that
the reverse containment holds, although the reverse 
containment is known to hold 
if $\exp$ only accesses its oracle nonadaptively.})
collapses to its
$\Delta_2$ level~\cite{hem:j:sky}.  
As to that brute-force bound,
for any typical, reasonable class $\cald$ of polynomial-size
circuits, and any reasonable complexity class $\calc$, the brute-force
bound one gets is that if $\calc$ has nonuniform circuits in $\cald$,
then $\calc$ has $\p^{\np^{\conp^{\calc}}}$-uniform circuits in
$\cald$ (equivalently, of course, has $\p^{\np^{\np^{\calc}}}$-uniform
circuits in $\cald$).\footnote{\label{foot:higher-and-lower}Recall that for example, the literature
  term P-uniform actually truly is speaking not of P but is speaking
  of FP, the polynomial-time computable functions, 
since we are speaking of which class is used to on input
  $1^j$ output an appropriate circuit to handle all inputs of length
  $j$, and that is a function issue.  But following that same
  convention, we speak of $\p^{\np^{\conp^{\calc}}}$-uniform, meaning
  that the function generating the circuits is actually in the class
  $\fp^{\np^{\conp^{\calc}}}$-uniform (and similarly, 
by $\pne$-uniform, we mean that the circuit generator
is in $\fp^{\NE}$).  We mention in passing that if
  one wants to make a slightly stronger claim, one can assert under
  the same assumption the existence of---here we really will name a
  function class, $\optp$~\cite{kre:j:optimization}, 
directly---$\optp^{\np^{\calc}}$-uniform circuits in
  $\cald$.  That slight improvement would not help us regarding
  Theorem~\ref{t:main} since an easy consequence of the collapse of
  the strong exponential hierarchy is that each $\optp^{\np^{\NE}}$
  function in fact is in $\fp^{\NE}$.

Eric Allender (personal communication, November 30, 2010)
interestingly pointed that one can go in the opposite direction,
namely, that the in general \emph{more} powerful class
FPSPACE$_{poly}^{\calc}$ (where the ``poly'' means that the 
FPSPACE machine asks only polynomially long queries to its 
oracle) can by brute-force cycle through circuits and inputs
and do the appropriate checking.  Although 
that class in general may be larger, and so one would not 
want to use this in such cases (in fact, similarly, 
the $\optp^{\calc}$ path noted above may well in some settings yield 
better claims than even
using $\fp^{\np^{\conp^{\calc}}}$), 
Eric notes that  
for the case 
$\calc = \NE$ it is not larger because the collapse of 
the strong exponential hierarchy has been extended
(see~\cite{sch-wag:c:collapsing,hem:j:census-space})
to
yield 
PSPACE$_{poly}^{\NE} = \pne$
(and even ${\rm NEXP}_{poly}^\NE = \pne$).}  This is true in the obvious
way.  The uniformity class is asserting that on input $1^j$ the 
$\fp^{\np^{\conp^{\calc}}}$ function outputs the circuit (description)
for the circuit that handles length-$j$ inputs.  Briefly put, this is
done as follows.  First, let $A$ be an arbitrary $\NE$ set that we
assume has (nonuniform) circuits in $\cald$.  Let $q$ be the polynomial
bound on the size of those circuits (or to be more precise, their
descriptions), and if they have a depth bound, let it be realized by
the constant-or-function $d$ (e.g., for $\acczero$, $d(j)$ 
would be any particular constant by which the hypothetical
$\acczero$ circuits for $A$ were depth-bounded).  
The $\fp$ part conducts a prefix search
to find the first 
(say, in dictionary ordering) 
$\cald$-type 
(i.e., within the allowed depth $d(j)$,
$q(j)$-size-bounded, and with the right types of gates and fan-in,
etc.)~circuit 
that is a correct circuit
for $A$ for all
length-$j$ inputs.  The $\np^{\conp^{\calc}}$ part supports this
in two ways (which way it is being used on a particular call will be
specified by an extra bit, not explicitly mentioned below, of
the call's argument string).  
It can answer the $\conp^{\calc}$ question: Here is
your input, which is $1^j$ and a circuit (description), and does that
described circuit correctly match $A$ in terms of acceptance/rejection
on all inputs of length $j$?  And it can also answer the
$\np^{\conp^{\calc}}$ question: Here is $1^j$ and a string $\alpha$ as
your input, and please let me know whether there exists a string
$\beta$ such that the combined length of $\alpha$ and $\beta$ is at
most $q(j)$ and the concatenation of $\alpha$ and $\beta$ is a circuit
(description) 
of $\cald$-type 
(i.e., is within the allowed depth $d(j)$,
$q(j)$-size-bounded, and uses the allowed types of gates and fan-in,
etc.)~that correctly matches $A$ in terms of
acceptance/rejection on all inputs of length $j$?  (Regarding this 
latter use, the $\np$ part is guessing the completion of the circuit,
the $\conp$ part is ranging over all length-$j$ strings, and for each
such string $y$ is seeing what the circuit does 
(we're assuming our circuits are such that with their description in
hand we can in polynomial time evaluate the circuit's action on a given 
string) on $y$
and then through one query to $\calc$ is finding whether 
$y \in A$, and then on that path of the $\conp$ machine accepts 
if the two actions agree---our $\conp$ machine model is that the 
machine by definition accepts exactly if all of its paths accept.)

For the particular case of $\acck$ and $\NE$, this gives that if $\NE$
has (nonuniform) $\acck$ circuits then there is a
$\fp^{\np^{\conp^{\NE}}}$ function that on input $1^j$ generates a
good circuit for length-$j$ inputs.  However, trivially,
$\fp^{\np^{\conp^{\NE}}} = \fp^{\np^{\np^{\NE}}}$, and since the
collapse of the strong exponential hierarchy~\cite{hem:j:sky} 
yields $\np^{\np^{\NE}} =
\p^{\NE}$, we have $\fp^{\np^{\conp^{\NE}}} = \fp^{\p^{\NE}} =
\fp^{\NE}$, i.e., we have $\pne$-uniformity.

\paragraph*{Acknowledgments}
All errors and blindnesses are the sole responsibility of the author,
who is not a specialist in circuit theory.  The author thanks Eric
Allender for the comments mentioned in
footnote~\ref{foot:higher-and-lower} and for helpful discussions,
Daniel \v{S}tefankovi\v{c} for a helpful conversation about Williams's
result, and Ryan Williams for pointing him to the reference~\cite{all:j:permanent-threshold}.

\bibliographystyle{alpha}

\end{document}